\documentclass[showpacs,floatfix,amsmath,
superscriptaddress,amssymb,twocolumn,prb]{revtex4}
\usepackage{latexsym}
\usepackage{graphicx}
\begin{document}
\title{Fourier Transformed Scanning Tunneling Peaks in the $d$-density wave phase}
\author{Cristina Bena}
\affiliation{Department of Physics, University of California
at Santa Barbara, Santa Barbara, CA 93106, USA}
\author{Sudip Chakravarty}
\affiliation{Department of Physics and Astronomy,
University of California Los Angeles, Los Angeles, California 90095-1547, USA}
\author{Jiangping Hu}
\affiliation{Department of Physics and Astronomy,
 University of California Los Angeles, Los Angeles, California 90095-1547, USA}
\author{Chetan Nayak}
\affiliation{Department of Physics and Astronomy,
University of California Los Angeles, Los Angeles, California 90095-1547, USA}
\date{\today}

\begin{abstract}
In this brief note we repeat an earlier calculation of
the Fourier transformed scanning tunneling spectra of
the $d$-density wave (DDW) phase using
a different band structure, which is more realistic and
consistent with the angle resolved photoemission spectroscopy (ARPES) data. We note that four peaks, which used to be located at $(\pm \pi/4,0)$ and  $(0, \pm \pi/4)$, are still 
present, but at positive energies their wavevectors shift to the neighborhood of
$(\pm 2 \pi/5,0)$, $(0, \pm 2\pi/5)$ and slowly disperse with energy.  The implications for the sensitivity with respect to the band structure are discussed.

\end{abstract}
\maketitle

Scanning tunneling microscopy (STM) experiments on the cuprate
high-temperature superconductors give unique information
about the local short-distance electronic structure of these
materials. Careful analysis of this information can yield
insights into the nature of the superconducting state 
\cite{hoffman1,hoffman2,howald}
(including, or perhaps especially, in the presence of vortices \cite{vortex})
and also into the short and meso-scale structure that
must develop as a precursor to superconductivity.
Recently, it has even been possible to perform such experiments
in the pseudogap regime above $T_c$ \cite{yazdani}  and also
to study their doping dependence in the superconducting state.\cite{sdavis} These experiments
give us a window on the gap (or depletion of low-energy states)
which is present in these materials and on its variation
from place to place in a given sample. One striking aspect of
all of these measurements is the presence of peaks in the
Fourier transform of the local electronic density of states (LDOS)
at wavevectors $(\pm 2\pi/\lambda,0)$, and $(0,\pm 2\pi/\lambda)$,
where the wavelength $\lambda$ is between $4$ and $7$
lattice spacings.\cite{sdavis} This problem has been analyzed from
a variety of theoretical perspectives
 \cite{Scalapino,Scalapino2,Dunghai,qp,c1,c2,c3,c4,c5,c6,Pereg-Barnea03,Misra04}.

One class of proposed explanations of the pseudogap state of the
cuprates is that it is due to spontaneous symmetry breaking
which occurs below the pseudogap scale ${T^*}(x)$ so long
as the doping is less than some critical value $x_c$
(and, presumably, larger than some minimum value).
According to one proposal \cite{ddw}, the broken-symmetry state
has $d$-density wave (DDW) order  and the pseudogap
appears to be `pseudo' only because this order parameter
is difficult to observe directly. Nevertheless, some elastic polarized neutron
scattering experiments \cite{Mook} seem to have observed this order directly in
underdoped YBCO (although other similar experiments,
but using unpolarized neutrons, have not found it \cite{Buyers}).
Furthermore, superfluid density, ARPES, Hall number, and
infrared Hall angle measurements \cite{otherexpddw,Chakravarty03}
which are indirect tests of the order
present in the pseudogap (if any) are consistent
with DDW order. Can STM measurements settle this question by
directly seeing the presence or absence of DDW order? Unfortunately, no.
In the DDW state, there is a lattice scale pattern of currents
which are staggered from one plaquette to another.
Tunneling, on the other hand, is sensitive to the local presence or
absence of charge available for tunneling. It is
not sensitive to currents unless they are accompanied by
charge excesses or deficits. Since DDW order breaks time reversal symmetry, defects
involving spin-orbit coupling can mix charge order, as ``angular momentum'' is not a
good symmetry in this case. 

However, STM measurements can be an indirect probe
of DDW order. Observable variations of the charge density can be induced by impurities.
They occur at wavevectors which connect points on
the contour in momentum space at which the
energy is equal to the applied voltage.
The peaks are strongest when the joint density of states of the
two ${\bf k}$-space points is greatest. According to this
picture, Fourier-transformed LDOS peaks
are due to the momentum-space structure
of the single-particle gap.
Such an explanation gives a compelling picture of the LDOS
peaks seen in the superconducting state 
\cite{hoffman2,Scalapino,Scalapino2,Dunghai}. As the voltage
is increased, the Fermi points expand into ovals which
stretch into banana-shaped loops. The corresponding
wavevectors seen in STM measurements roughly follow the
evolution of these energy contours. In the pseudogap state, however,
the LDOS peaks hardly disperse with energy \cite{yazdani}.
Is this indicative of some entirely different phenomenon which
is causing the peaks in the pseudogap regime?

\begin{figure*}
\begin{center}
\includegraphics[width=5in]{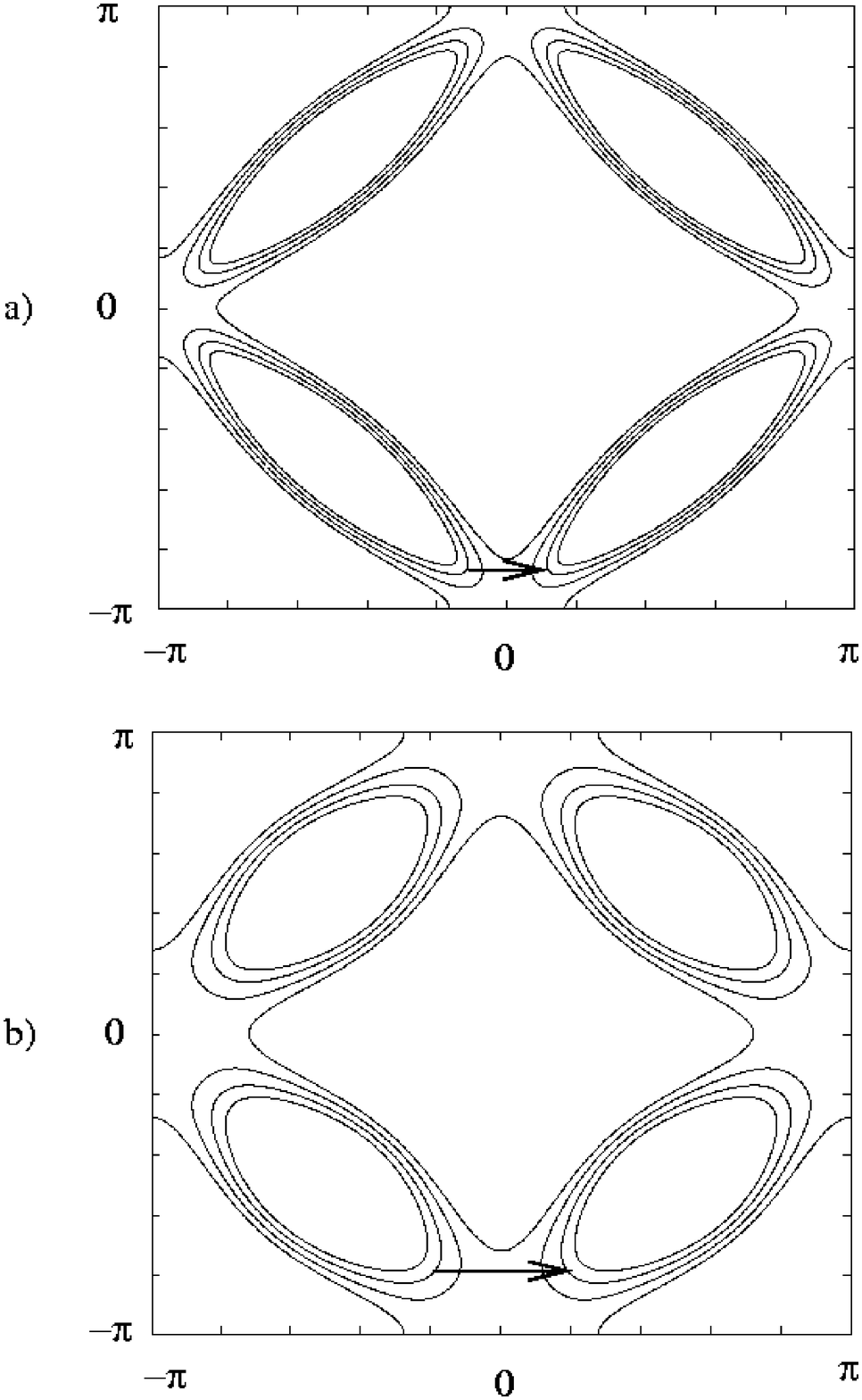}
\end{center}
\caption{Equal energy contours for the pure DDW state with a) $t-t'$ band
structure b) ARPES consistent
band structure. The magnitude of the DDW gap is taken to be $40 \rm
meV$.The innermost contour corresponds to $+40 \rm meV$
energy, while the outermost contour corresponds to a $-40 \rm meV$
energy. Scattering between regions of high density of states is indicated
by the arrows. }
\label{bs}
\end{figure*}

In this note, we argue that it is not. In a recent paper\cite{qp},
we examined the patterns resulting 
from quasiparticle scattering in the Fourier transform
of the local density of states for a high $T_c$ superconductor for which
we assumed that the pseudogap is described by $d$-density wave order.
We considered a $t-t'$ band structure, and in the spectra we found peaks
in the LDOS centered about $(\pm \pi/4,0)$, $(0, \pm \pi/4)$, which dispersed 
slowly with  energy. Here, we show that their
precise $k$-space location and dispersion with energy are dependent
on the details of the band structure.  In particular, if we use
a more realistic band structure than we used previously,
we find LDOS peaks at positive voltage bias which are located closer to where
they have been observed experimentally -- in the vicinity of
$(\pm 2 \pi/5,0)$, $(0, \pm 2\pi/5)$ (although their location varies
with doping \cite{sdavis})
as opposed to the wavevectors $(\pm \pi/4,0)$, $(0, \pm \pi/4)$ with the previous
band structure -- and which disperse less. The formalism and the method of calculation are identical to those in Ref.~\onlinecite{qp}. So, we shall not repeat them here. The band structure which
we use is flatter near the $(\pi,0)$ point, consistent with
ARPES measurements  \cite{norman}. It is characterized by the
energy dispersion (with the lattice spacing set to unity)
\begin{multline}
\epsilon_k={t_0} +{t_1}(\cos k_x+\cos k_y)/2 +
{t_2}\cos k_x \cos k_y\\
 + {t_3}(\cos 2 k_x + \cos 2 k_y)/2+{t_5}\cos 2 k_x \cos 2 k_y\\
+ {t_4}(\cos 2 k_x \cos k_y+ \cos 2 k_y \cos k_x)/2\\
\end{multline}
and $t_{0-5}=0.1305,-0.5951,0.1636,-0.0519,-0.1117$, $0.0510 ({\rm eV})$.
We also take the chemical potential shift (from the above dispersion)
to be $\delta\mu=-0.034 {\rm eV}$.
The chemical potential is chosen such that, consistent with the ARPES measurements
\cite{norman},  no electron pockets open in the band structure.
The equal energy contours for this ARPES band structure as well as for the
simpler band structure used in ref. \onlinecite{qp} in which
${t_3}, {t_4},{t_5}$ vanish are given for comparison in Fig. \ref{bs}.
We focus on the case of non-magnetic impurity scattering. 

\begin{figure*}
\begin{center}
\includegraphics[width=6in]{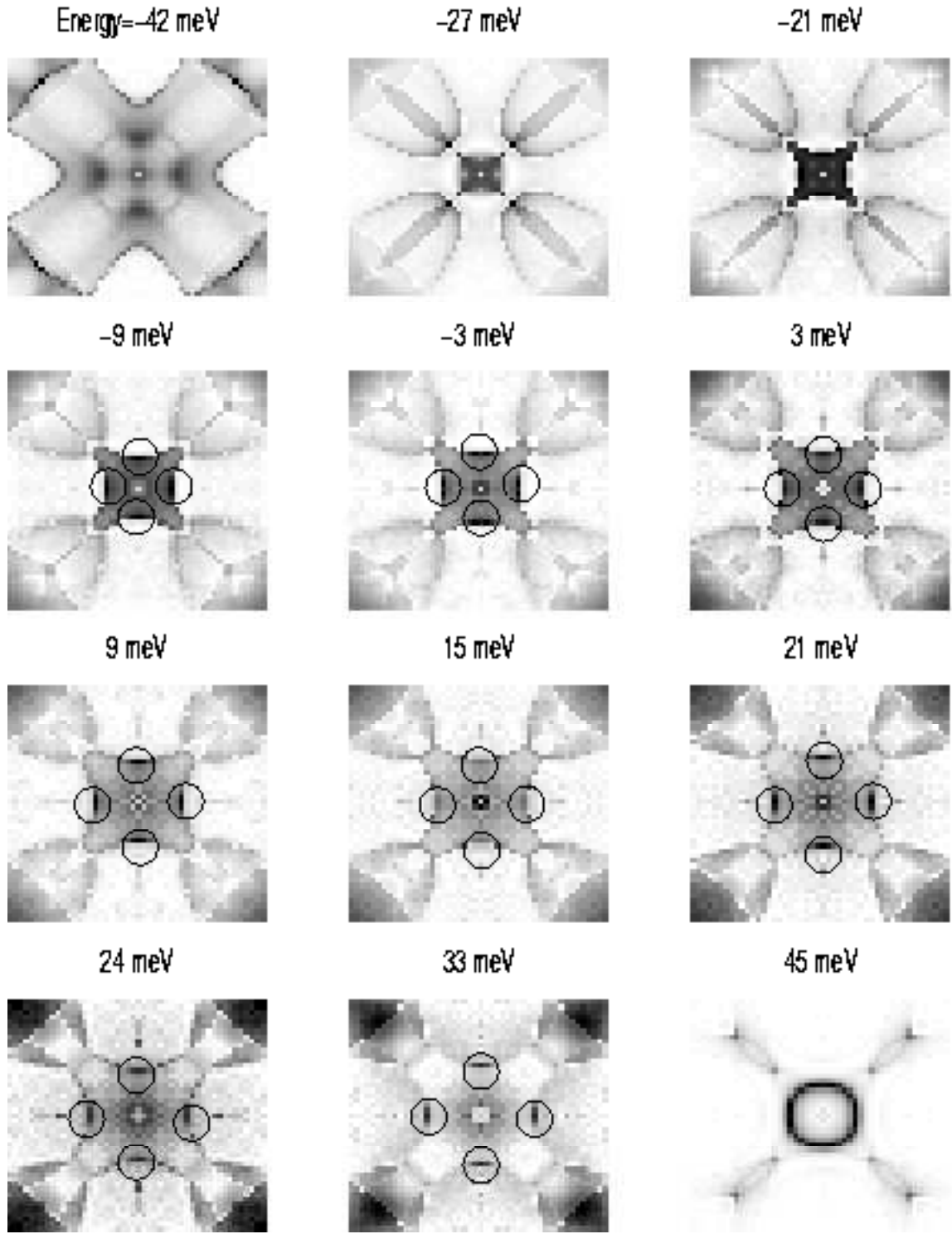}
\end{center}
\caption{Quasiparticle interference spectra for a DDW state with DDW gap
of $40 \rm meV$ for non-magnetic impurity scattering $V=0.1 \rm eV$. The
results are displayed for energies ranging from $-42 \rm meV$ to $45 \rm
meV$ on a linear gray scale. Each plot represents the spectral intensity
as a function of momentum in the 1st BZ, for $|q_x|, |q_y|<\pi$. The
positions of the peaks are indicated by circles}
\label{res}
\end{figure*}

Our results are plotted in Fig. \ref{res} for energies between
$-42 {\rm meV}$ and
$45 {\rm mev}$ for a pure DDW state with a gap of $W_0=40 {\rm meV}$.
These calculations are identical to those of Ref.~\onlinecite{qp}.
We note the appearance of peaks corresponding to
scattering between the tips of the ellipses, as indicated by arrows in
Fig. \ref{bs}. The position of the peaks is marked by circles in Fig. 
\ref{res}.

\begin{figure*}
\begin{center}
\includegraphics[width=5in]{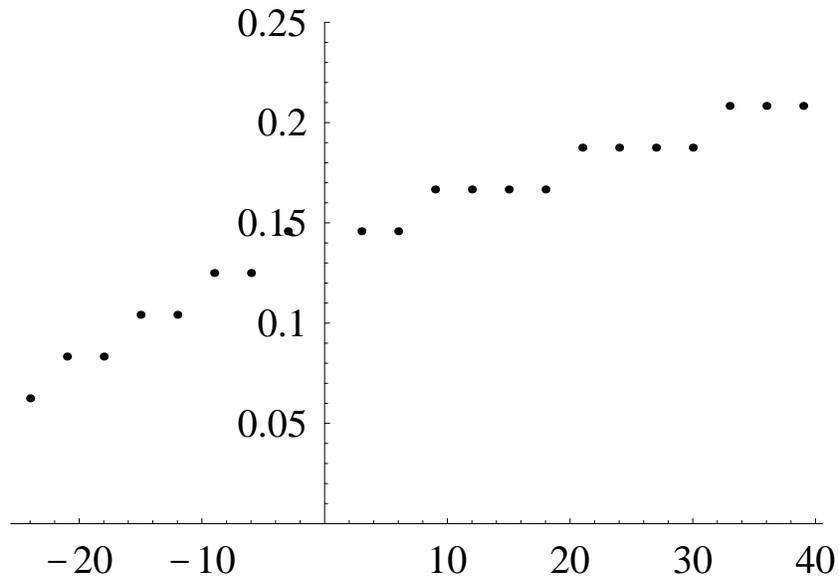}
\end{center}
\caption{The energy dispersion of the wavevectors of the observed peaks. 
The magnitude of the wavevectors is plotted in units of $2 \pi$, and the 
energy is measured in $meV$. The magnitude of the $DDW$ gap is $40 {\rm 
meV}$.}
\label{dis}
\end{figure*}

The dispersion of the wavevectors with energy is plotted in 
Fig.\ref{dis}. As indicated, for positive energies, the dispersion
is quite small, and the magnitude of the wavevectors ranges from  $(2\pi)/6.9$
at $3 {\rm meV}$ to  $(2 \pi)/4.8$ at $39 {\rm meV}$. The dispersion
is larger for  negative energies, and the peaks are not as well-defined. 
We also note that if one does not shift the chemical potential, while electron pockets open, the
dispersion of the peaks position with energy is much smaller and thus more in agreement with the
experimental results. This indicates that both the position of the peaks and their energy
dispersion are very sensitive to band structure parameters.

Thus, the position of the peaks resulting from quasiparticle interference
in a pure DDW with an ARPES-consistent band structure is
similar to experimental observations,
while the dispersion with energy is larger. However, this 
is a crude single impurity scattering T-matrix
approximation calculation  and other factors may need to be taken into
account. Electron-electron interactions have been neglected, but are
likely to be important for understanding ARPES experiments
\cite{Chakravarty03} in which they smear out antinodal quasiparticles.
It is hard to imagine that this would not have an impact on LDOS
measurements. Furthermore, the renormalization of the order parameter due to
disorder in the presence of many impurities may need to be taken into
account through a self consistent Bogoliubov-de Gennes calculation\cite{Ghosal03}.

In conclusion, we obtained the quasiparticle interference spectra in a DDW
state, which would correspond to the pseudogap phase of the cuprates. We
used an ARPES consistent band structure. We observed the emergence of
peaks with wavevectors of magnitude $2 \pi/4 - 2\pi/7$(at positive 
energies) along the ($\pi,0$)
direction. The magnitude of the wavevectors is similar to those seen in recent STM
experiments in the pseudogap phase \cite{yazdani,sdavis,vortex}. 
The energy dispersion of the peaks is larger than what was observed experimentally,
though other theoretical and experimental factors may need to be taken
into account when trying to connect our observations to the experimental
data. However, the peaks seen in STM experiments will always
disperse with energy -- even in the case of charge order disrupted
by impurities \cite{c1,c2}.
The issue at hand is the quantitative one of how much.

\acknowledgments
C. B. has been supported by the NSF
under Grant No. DMR-9985255, and also by funds from the A. P.
Sloan Foundation and the Packard Foundation. S. C. has been supported by the NSF
under Grant No. DMR-9971138.
J. P. Hu was supported by the funds from
the David Saxon chair at UCLA.
C. N. has been supported by the NSF under Grant No. DMR-9983544
and the A.P. Sloan Foundation. We would like to thank S. Kivelson for helpful discussions.

\end{document}